\documentclass{PoS}

\title{Exploring the squark flavour structure of the MSSM}
\ShortTitle{Exploring the squark flavour structure of the MSSM}

\author{Karen De Causmaecker\\
        Theoretische Natuurkunde, IIHE/ELEM and International Solvay Institutes, Vrije Universiteit Brussel, Pleinlaan 2, 1050 Brussels, Belgium}

\author{Benjamin Fuks\\
        Sorbonne Universit\'es, UPMC Univ.\ Paris 06, UMR 7589, LPTHE, 75005 Paris, France; CNRS, UMR 7589, LPTHE, 75005 Paris, France; Institut Universitaire de France, 103 boulevard Saint-Michel, 75005 Paris, France }

\author{\speaker{Bj\"orn Herrmann}\\
		LAPTh, Universit\'e Savoie Mont Blanc, CNRS, 9 Chemin de Bellevue, 74941 Annecy-le-Vieux, France}

\author{Farvah Mahmoudi\\
		Univ Lyon, Univ Lyon 1, ENS de Lyon, CNRS, Centre de Recherche Astrophysique de Lyon UMR 5574, F-69230 Saint-Genis-Laval, France;
		Theoretical Physics Department, CERN, CH-1211 Geneva 23, Switzerland}

\author{Ben O'Leary, Werner Porod\\
		Institut f\"ur Theoretische Physik und Astrophysik, Universit\"at W\"urzburg, 97074 W\"urzburg, Germany}

\author{Sezen Sekmen\\
        Kyungpook National University, Department of Physics, Daegu, 702-701 Korea }

\author{Nadja Strobbe\\
        Fermi National Accelerator Laboratory, Batavia, 60510-5011, USA }

\abstract{We present an extensive study of the MSSM parameter space allowing for general generation mixing in the squark sector. Employing an MCMC algorithm, we establish the parameter ranges which are allowed with respect to various experimental and theoretical constraints. Based on this analysis, we propose benchmark scenarios for future studies. Moreover, we discuss aspects of signatures at the LHC.}

\FullConference{38th International Conference on High Energy Physics\\
		3-10 August 2016\\
		Chicago, USA}

\begin{document}

% =================================================
\section{Introduction}

In the Standard Model, the Yukawa matrices are the only source of flavour-violation, leading to quark flavour-violating interactions related to the CKM-matrix. Considering new physics, there are two ways of modeling the flavour structure of the theory. Either we assume the same flavour structure as in the Standard Model, i.e.\ all flavour-changing currents remain related to the CKM-matrix, or we allow for new sources of flavour violation. The first scenario is called Minimal Flavour Violation (MFV), while the latter case is referred to as Non-Minimal Flavour Violation (NMFV).

In the squark sector of the MSSM, non-minimally flavour violating terms manifest as non-diagonal entries in the soft-mass matrices $M^2_{\tilde{Q}}$, $M^2_{\tilde{U}}$, and $M^2_{\tilde{D}}$, which are related to left-handed squarks, right-handed up-type squarks, and right-handed down-type squarks, respectively. Moreover, off-diagonal NMFV entries can be present in the trilinear coupling matrices $T_u$ and $T_d$ related to up- and down-type squarks, respectively. It is convenient to parametrize NMFV in terms of seven dimensionless variables, 
\begin{equation}
	\delta_{LL} = \frac{ \big( M^2_{\tilde{Q} }\big)_{23} }{ \big( M_{\tilde{Q} }\big)_{22} \big( M_{\tilde{Q} }\big)_{33}}, \quad 
	\delta^u_{RR} = \frac{ \big( M^2_{\tilde{U} }\big)_{23} }{ \big( M_{\tilde{U} }\big)_{22} \big( M_{\tilde{U} }\big)_{33}}, \quad
	\delta^u_{RL} = \frac{v_u}{\sqrt{2}} \frac{ \big( T_u \big)_{23} }{ \big( M_{\tilde{U} }\big)_{22} \big( M_{\tilde{Q} }\big)_{33}},
\end{equation}
and so on. Here, $v_u$ denotes the vacuum expectation value of the Higgs doublet coupling to up-type (s)quarks. Note that we only consider additional flavour mixing between the second and third generation of squarks, which is experimentally less constrained and phenomenologically most interesting. 

While previous studies of NMFV in the squark sector were mainly based on only one of the NMFV parameters defined above being non-zero, the aim of the present analysis is to study the more general situation, where all flavour-violating entries of the Lagrangian are potentially sizeable. In the following we briefly outline the setup and main results of this study. A more detailed discussion can be found in Ref.\ \cite{JHEP2015}.

% =================================================
\section{An MCMC parameter study}

Our analysis concerns a 22-dimensional parameter space, involving the most important MSSM parameters (including the seven NMFV parameters defined above) related to the squark, gaugino, and Higgs sectors. In order to efficiently scan such a large parameter space, we employ the Markov Chain Monte Carlo (MCMC) technique. We impose a variety of theoretical and experimental constraints in order to identify the allowed regions of parameter space. Moreover, the MCMC technique leads to posterior distributions for each input parameter as well as for the physical mass spectrum. The experimental constraints are mainly from precision measurements in the sector of $B$- and $K$-mesons, see Ref.\ \cite{JHEP2015} for details. 

\begin{figure}[t]
	\begin{center}
		~~~~~\includegraphics[width=0.9\textwidth]{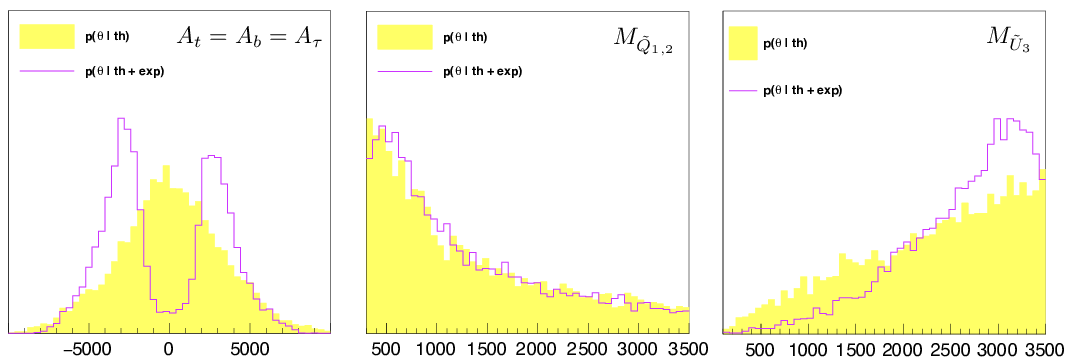}
	\end{center}
	\vspace*{-5mm}
	\caption{Prior (yellow) and posterior (purple) probability distributions of three selected flavour-conserving parameters of our NMFV-MSSM setup.}
	\label{Fig1}
\end{figure}

In Fig.\ \ref{Fig1}, we show the obtained distributions of three selected flavour-conserving parameters affecting the squark sector. The mentioned constraints strongly influence the shape of the distribution of the trlinear coupling featuring two peaks around $A_t \sim 3000$ GeV. As in the MFV case, this is explained by the Higgs-boson mass requiring a relatively large mass splitting between the squarks exhibiting the largest stop components, which is realized only for particular parameter combinations. 

\begin{figure}[t]
	\begin{center}
		\includegraphics[width=0.9\textwidth]{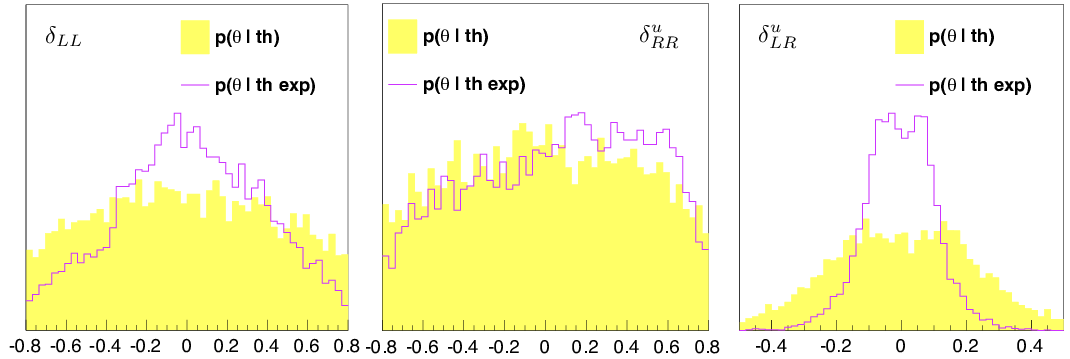}
	\end{center}
	\vspace*{-5mm}
	\caption{Prior (yellow) and posterior (purple) probability distributions of three selected flavour-violating parameters of our NMFV-MSSM setup.}
	\label{Fig2}
\end{figure}

An interesting and somewhat unexpected feature concerns the squark soft mass parameters. As can be seen in Fig.\ \ref{Fig1}, smaller values are preferred for the first and second generations, while the third generation prefers heavier masses. This is also mostly caused by imposing the Higgs-boson mass. The NMFV contributions to the latter involve the off-diagonal elements of the trilinear coupling matrice together with the diagonal entries of the soft mass matrices. The soft mass parameters associated to the third generation are pushed to relatively large values (see the example of $M_{\tilde{U},3}$ in Fig.\ \ref{Fig1}) due to the imposed constraints coming from the flavour sector. For a non-zero off-diagonal element of $T_u$, the Higgs becomes tachyonic if the mass parameter $M_{\tilde{Q}_{1,2}}$ associated to the first and second generations becomes too large. Moreover, requiring physical solutions or the electroweak vacuum also favour lower values of these parameters. 

Coming to the flavour-violating parameters, our study shows that most of the seven parameters can be sizeable in wide regions of the allowed parameter space. In Fig.\ \ref{Fig2} we show three examples of distributions. In addition to the Higgs boson mass, the most important constraints are found to be the meson-oscillation parameter $\Delta M_{B_s}$ and the decay $B_s \to \mu\mu$. The two peaks observed in the distribution of $\delta^u_{LR}$ are explained by the corrections to the Higgs-boson mass discussed above. In the sector of down-type squarks, the distributions of $\delta^d_{LR}$ and $\delta^d_{RL}$ show clear peaks around zero as large values mostly lead to tachyons, but are hardly constrained from the imposed flavour observables. 

Concerning the overall distribution of NMFV entries present in our model, we find that for essentially all scenarios at least one NMFV parameter is non-vanishing and sizeable, while at least one NMFV parameter has to be small. However, a large fraction of the scanned parameter points exhibit several non-vanishing NMFV elements. Let us finally mention that we did not find any strong correlations between the different NMFV parameters under consideration.

% =================================================
\section{LHC phenomenology and outlook}

%In order to represent the overall distribution of NMFV entries present in model under consideration, we introduce the quantities $\big| \vec{\delta} \big|$ and $\log \big| \Pi_{\delta}\big|$, which are defined in Fig.\ \ref{Fig3}. As can be seen in their respective distributions, for essentially all scenarios at least one NMFV parameter is non-vanishing and sizeable, while at least one NMFV parameter has to be small. However, a large fraction of the scanned parameter points exhibit several non-vanishing NMFV elements. Let us finally mention that we did not find any strong correlations between the different NMFV parameters under consideration.
%\begin{figure}[t]
%	\begin{center}
%		\includegraphics[width=0.6\textwidth]{Fig3.pdf}
%	\end{center}
%	\vspace*{-5mm}
%	\caption{Prior (yellow) and posterior (purple) probability distributions of the two indicators illustrating the global distribution of NMFV in our setup.}
%	\label{Fig3}
%\end{figure}

\begin{figure}[t]
	\begin{center}
		\includegraphics[width=0.9\textwidth]{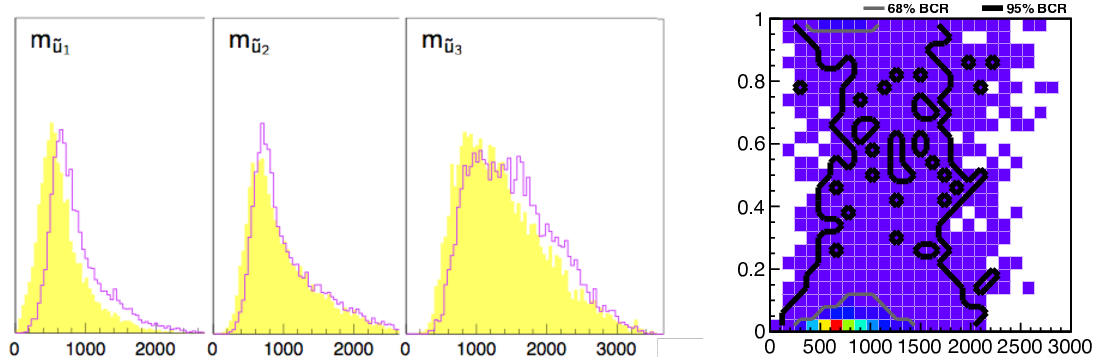}
	\end{center}
	\vspace*{-5mm}
	\caption{Left: Prior (yellow) and posterior (purple) probability distributions of the masses of the three lightest up-type squarks. Right: Correlation between the mass and the stop-content of the lightest up-type squark.}
	\label{Fig4}
\end{figure}

Coming to the physical masses of the squarks, selected distributions are shown in Fig.\ \ref{Fig4}. Our analysis shows that, despite the strong constraints on the parameter space, the three lightest up-type squarks are likely to have masses around 1 TeV or less, and should therefore be accessible at the LHC. As can be seen in Fig.\ \ref{Fig4}, we find that the lightest up-type squark is likely to be rather charm-like than stop-like. This is in contrast to the MFV-MSSM and is a direct consequence of the fact that the distributions of the mass parameters of the first and second generations prefer lower values, while the distributions related to the third generation are pushed towards higher mass values. 

Finally, based on the results of the complete MCMC study, we propose four benchmark scenarios within the MSSM with non-minimal flavour violation in the squark sector. These four scenarios capture typical features, such as flavour decompositions and branching fractions, identified in the MCMC analysis, and are destinated for future studies of NMFV at the LHC. 

All details on our analysis, the conclusions on the sectors of up- and down-type squarks, as well as the defined benchmark points can be found in Ref.\ \cite{JHEP2015}.

% =================================================

\end{document}